\documentclass[epj]{webofc}
\usepackage[varg]{txfonts}   
\wocname{EPJ Web of Conferences}
\woctitle{CONF12}

\usepackage{amsmath,amssymb}
\usepackage{dsfont}
\usepackage{slashed}
\usepackage{verbatim}
\usepackage{graphicx}

\usepackage{placeins}

\newcommand{\cl}{\centerline}
\newcommand{\be}{\begin{equation}}
\newcommand{\ee}{\end{equation}}
\newcommand{\bea}{\begin{eqnarray}}
\newcommand{\eea}{\end{eqnarray}}

\begin{document}
\selectlanguage{english}
\title{\cl{New Physics Searches from } \cl{Nucleon Matrix Elements in Lattice QCD}}
%
%

\author{Martha Constantinou\inst{1}\fnsep\thanks{\email{marthac@temple.edu}}}

\institute{Temple University}

\abstract{
In this paper we review recent progress in hadron structure using
lattice QCD simulations, with main focus in the evaluation of nucleon
matrix elements. We highlight developments that may guide new
Physics searches, such as the scalar and tensor charges, as well as, the 
neutron electric dipole moment.
}
\maketitle
\section{Introduction}
\label{sec1}

For more than a decade hadron structure has been at the forefront of Nuclear and Particle Physics research, with the proton being an 
ideal laboratory for studying the QCD dynamics. However, after almost half a century since the first exploration of the internal structure 
of the proton at SLAC ~\cite{Breidenbach:1969kd,Bloom:1969kc} there are still open questions, such as the electromagnetic charged radii, 
and the spin distribution among the quarks and gluons. In theoretical hadronic Physics, the main challenge 
is to provide a quantitative description of strongly interacting particles using the underlying theory of QCD. The only ab initio formulation is 
Lattice QCD (LQCD), that enables one to study the properties of fundamental particles numerically. This is done by defining the continuous 
equations on a discrete Euclidean four-dimensional lattice, which results in equations with hundreds of billions of variables, and must be simulated 
in powerful computers. Moreover, at the hadronic energy scale where perturbative methods cannot be applied to solve QCD, the non-perturbative 
formulation of LQCD is of the utmost importance.

The nucleon, one of the building-blocks in the universe, is the only stable hadron in the Standard Model, yet its structure is not fully understood. This has been a major goal of LQCD simulations, and over the last years there has been a striking progress in this direction, due to algorithmic improvements, theoretical developments and increase of computational resources. The latest achievement of the field of LQCD is simulations at
parameters tuned to their physical values (see e.g., Refs.~\cite{Green:2012ud,Green:2014xba,Abdel-Rehim:2015owa,Yang:2015uis,Bhattacharya:2016zcn} and the 
reviews of Refs.~\cite{SaraLat16,ZanottiLat15,Constantinou:2014tga,Syritsyn:2014saa,Lin:2012ev}). We will refer to such simulations as:
simulations at the ``physical point''. This new era of LQCD calculations can provide high-precision input to on-going and planned 
experiments, test phenomenological models, and predict Physics beyond the Standard Model (BSM). 

In these proceedings we discuss representative observables probing nucleon structure, focusing on quantities that are not easily accessible 
experimentally, and are probes of Physics BSM, or related to direct searches of dark matter. More precisely, we discuss recent developments 
in calculations of the scalar and tensor charge, and the neutron electric dipole moment (nEDM).

\section{Setup}
\label{sec2}

\begin{figure}[!h]
\cl{\includegraphics[scale=0.6]{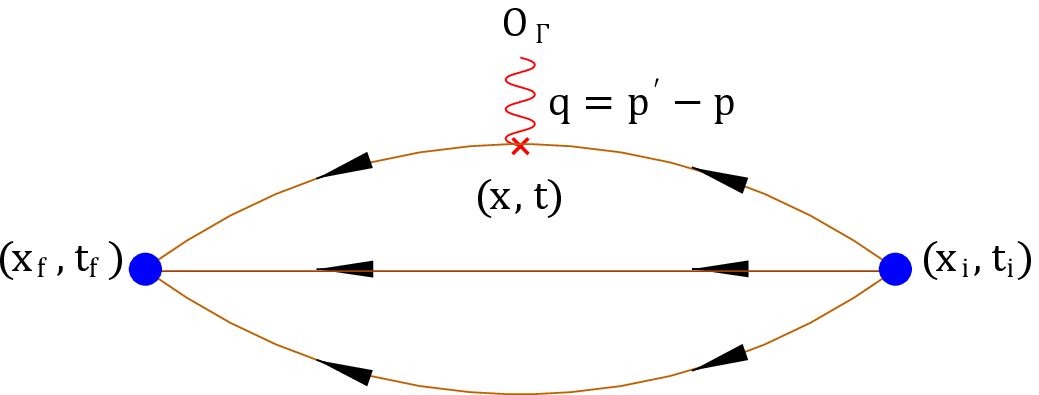} \hspace{.8cm}
\includegraphics[scale=0.6]{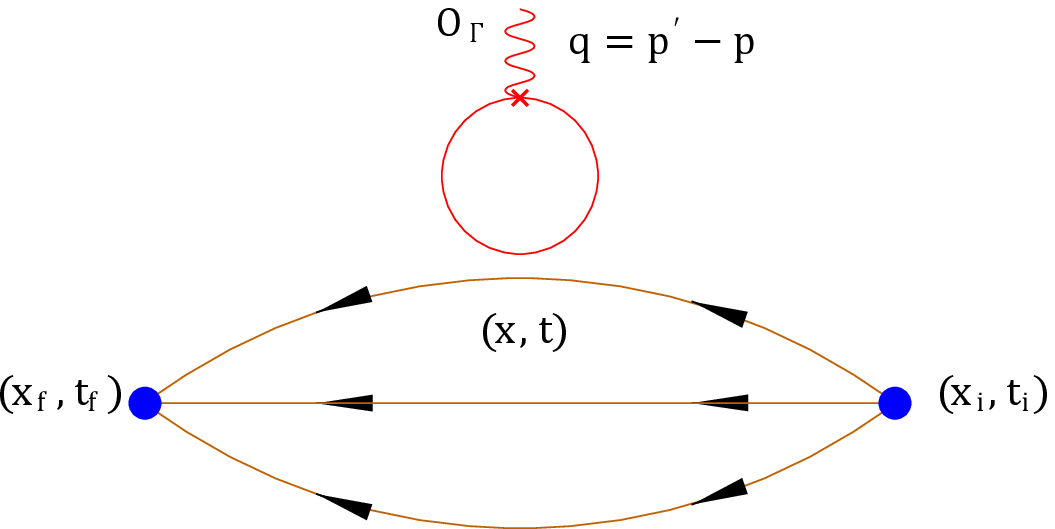}}
\vskip 0.25cm
\caption{Connected (left) and disconnected (right) quark contributions to
the nucleon three-point function.}
\label{fig1}
\end{figure}

There are two types of diagrams (Fig.~\ref{fig1}) that enter the evaluation of the quark contributions related to nucleon structure within LQCD, 
the connected (left) and disconnected (right). The disconnected diagram has been neglected until recently because it is very noisy 
and expensive to compute. During the last few years a number of groups are studying various techniques for its computation
\cite{Babich:2010at,Engelhardt:2012gd,Stathopoulos:2013aci,Abdel-Rehim:2013wlz,Chambers:2015bka,Gambhir:2016jul} and results at the
physical point already appear in the literature \cite{Yang:2015uis,Abdel-Rehim:2016won,Abdel-Rehim:2016pjw} (see Ref.~\cite{SaraLat16} 
for a recent review).
\vskip -0.4cm
\FloatBarrier
In the computation of nucleon matrix elements one needs  
appropriate two- and  three-point correlation functions defined as: 
\bea
\hspace*{2.5cm}
 G^{2pt}(\vec q, t_f) &=& \sum_{\vec x_f} \, e^{-i\vec x_f \cdot \vec q}\,
\Gamma^0_{\beta\alpha}\, \langle {{J_{\alpha}(\vec x_f,t_f)}}{{\overline{J}_{\beta}(0)}} \rangle \,, \\
 G^{3pt}_{\cal O}(\Gamma^\mu,\vec q, t_f) &=& \sum_{\vec x_f, \vec x} \, e^{i\vec x \cdot \vec q}\,e^{-i\vec x_f \cdot \vec p'}
     \Gamma^\mu_{\beta\alpha}\, \langle { J_{\alpha}(\vec x_f,t_f)} {\cal O}(\vec x,t) {\overline{J}_{\beta}(0)} \rangle\,,
\eea
\noindent
where the projectors $\Gamma^\mu$ are usually defined as
$\Gamma^0 \equiv \frac{1}{4}(1+\gamma_0),\,\, \Gamma^k \equiv \Gamma^0\cdot\gamma_5\cdot\gamma_k\,$ in order to evaluate
the quantities of interest. The lattice data are extracted from dimensionless ratios of the two- and three-point correlation functions:
\be
R_{\cal O}(\Gamma,\vec q, t, t_f) = \frac{G^{3pt}_{\cal O}(\Gamma,\vec q,t)}{G^{2pt}(\vec 0, t_f)}
\hspace*{-0.1cm}\times\hspace*{-0.1cm}\sqrt{\frac{G^{2pt}({-}\vec q, t_f{-}t)G^{2pt}(\vec 0, t)G^{2pt}(\vec 0, t_f)}{G^{2pt}(\vec 0  , t_f{-}t)G^{2pt}({-}\vec q,t)G^{2pt}({-}\vec q,t_f)}}\,\,
{\rightarrow \atop {{t_f{-}t\rightarrow \infty} \atop {t{-}t_i\rightarrow \infty}}}\,\,
\Pi (\Gamma,\vec q) \,,
\label{EqRatio}
\ee
which is considered optimized since it does not contain potentially noisy two-point functions at large time separations 
and because correlations between its different factors reduce the statistical noise. The most common method to extract the 
desired matrix element is to look for a plateau with respect to the current insertion time, $t$. 
The current should couple to the nucleon at a time $t$, that is away from the source and the sink timeslices in order
to ensure single state dominance. Alternative analysis methods for the investigation of excited states contamination  
is a two- or three-state fits to take into account contributions from excited states, as well as, the summation method, in which one sums 
the ratio from the source to the sink (one may exclude the source and sink points to avoid contact terms). In the latter method 
the excited state contaminations are suppressed by exponentials decaying with $(t_f- t_i)$. 

For a meaningful comparison with experimental data and phenomenological estimates, one needs to apply 
renormalization on the matrix elements\footnote{unless a conserved current is used.} which, for the quantities discussed in 
this review, is multiplicative:
\be
\Pi^R (\Gamma,\vec q) = Z_{\cal O}\,\Pi (\Gamma,\vec q)\,.
\ee
The renormalizes matrix elements may be parameterized in terms of
Generalized Form Factors, following the symmetry properties of QCD. In these proceedings 
we will mostly discuss about nucleon charges, which are the forward limit of nucleon matrix elements,
that is, there is zero momentum transfer ($Q^2{=}0$) between the initial and final states.

\section{Results}
\label{sec3}

The class of observables that are computed in lattice simulations related to nucleon structure are two kind: {\bf 1.} those that are 
well-known experimentally, and thus, LQCD results provide postdictions, and {\bf 2.} those that are lesser-known or difficult to measure, 
and lattice simulations may give valuable predictions. Quantities of the former type is the axial charge and electromagnetic form factors, and serve 
as a benchmark of the LQCS formalism. Reproducing experimental data from first principle calculations, will give confidence in providing
predictions for quantities that can impact new Physics searches, such as the scalar $\&$ tensor charges and the nEDM, discussed below.

\subsection{Scalar $\&$ Tensor Interactions}
\label{sub3_1}

It is only recently that the tensor and scalar charges have been studied in LQCD with quantified systematics, mainly due to the fact that the 
contributions of effective scalar and tensor interactions are very small (per-mil level). These interactions correspond to the non $V-A$ 
structure of weak interactions and serve as a test for new Physics.  
However, ongoing experiments using ultra-cold neutrons \cite{Abele:2002wc,Nico:2009zua,Young:2014mxa,Baessler:2014gia}, 
as well as planned ones~\cite{Bhattacharya:2011qm} will reach the necessary precision to investigate such interactions, increasing thus
the interest for results from {\it ab initio} calculations. 

In order to study the scalar and tensor interactions we add a term in
the effective Hamiltonian corresponding to new BSM Physics, 
\be
H_{eff} = G_F \left( J_{V_A}^{l}\times J_{V_A}^{q} + \sum_i \epsilon_i {\cal O}_i^l \times {\cal O}_i^q \right)
\ee
which includes operators with novel structure, such as the
scalar and tensor. These come with low-energy couplings that are
related to masses of new TeV-scale particles. 

\subsubsection{Scalar Charge}
\label{sub3_1_1}

Computations of the scalar charge may provide input for dark matter searches, since direct-detection experiments 
are based on measuring the recoil energy of a nucleon hit by a dark matter candidate (WIMP). In supersymmetric scenarios
\cite{Ellis:2010kf} the interaction between the nucleon and dark matter is mediated through a Higgs boson and the
spin independent scattering amplitude at $Q^2{=}0$ involves the quark content of the nucleon ($\sigma$-term), which is closely 
related to the scalar charge. There are  a few calculation on the $\sigma$-terms that will not be presented here
due to space limitations, and more details can be found in Refs.~\cite{Yang:2015uis,Abdel-Rehim:2016won,Bali:2016lvx}.

To extract the scalar charge we compute the nucleon matrix element
\be
\langle N(\vec{p'}) \, \mathcal{O}_X \, N(\vec{p}) \rangle \Big{|}_{Q^2=0} \,, \quad 
\mathcal{O}^\alpha_X = \mathcal{O}^\alpha_S \equiv \overline{\Psi}(x)\,\frac{\tau^\alpha}{2}\, \Psi(x)\,.
\ee
In order to provide individual quark contributions one needs to compute both the isovector ($\alpha{=}3$) and
isoscalar ($\alpha{=}1$) combinations, $u{-}d$ and $u{+}d$, respectively. While for the isovector combination the disconnected
contribution cancels out, the isoscalar receives contributions. The disconnected contributions are not negligible for the case of the 
scalar current and, thus, cannot be ignored in the contribution from individual quarks and isoscalar flavor combination, as it would 
lead to unreliable estimate for the scalar charge and the nucleon $\sigma$-terms. The latter have
been studied recently for the up, down, strange and charm quarks~\cite{Yang:2015uis,Abdel-Rehim:2016won,Bali:2016lvx}.

There are limited studies on the scalar charge due to the fact that it has the smallest signal-to-noise ratio as
compared to the other nucleon charges. In particular, at least an order of magnitude more statistics is required
to achieve the same accuracy as the axial and tensor charges. In addition, it exhibits severe contamination from 
excited states and, thus, the systematic uncertainties are not well-controlled. Moreover, the scalar current couples 
to the vacuum, and thus, a vacuum subtraction is required. 
\begin{figure}[h]
\cl{\includegraphics[scale=0.32,angle=-90]{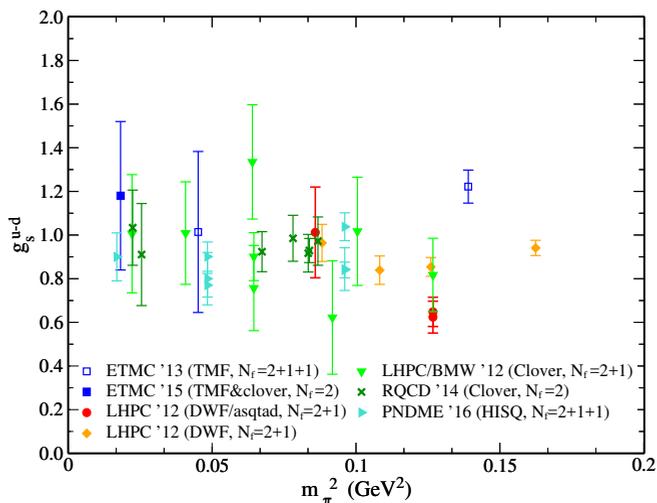}}
\caption{Lattice results for $g^{u-d}_S$ as a function of $m_\pi^2$, corresponding to: 
$N_f{=}2{+}1$ Clover fermions by LHPC/BMW~\cite{Green:2012ej} (down green triangles);
$N_f{=}2{+}1$ DWF by LHPC~\cite{Green:2012ej} (orange diamonds);
$N_f{=}2{+}1$  DWF/staggered by LHPC~\cite{Green:2012ej} (red circles);
$N_f{=}2{+}1{+}1$ TMF by ETMC~\cite{Alexandrou:2013jsa} (open blue square);
$N_f{=}2$ Clover fermions by RQCD~\cite{Bali:2014nma} (green x);
$N_f{=}2$ TMF $\&$ Clover by ETMC~\cite{Abdel-Rehim:2015owa} (filled blue square);
$N_f{=}2{+}1{+}1$ HISQ by PNDME~\cite{Bhattacharya:2016zcn} (turquoise right triangles).} 
\label{fig_gS}
\end{figure}
\FloatBarrier
In Fig.~\ref{fig_gS} we show results for isovector scalar charge, $g^{u-d}_S$, as a function of the pion mass squared, $m_\pi^2$. These have been obtained by 
various groups using different discretizations, lattice volumes, number of dynamical quarks and values for the lattice 
spacing. The results presented are renormalized using the $\overline{MS}$-scheme and evolved at a scale of 2 GeV.
The continuum limit has to be taken before comparing the results from different discretizations, however, not all data are available at
multiple values of the lattice spacing. Nevertheless, a comparison between different collaborations shows an overall good agreement among all 
lattice results obtained with similar source-sink separation, which is only qualitative due to the large statistical errors. The results plotted in Fig.~\ref{fig_gS} 
correspond to: 
$N_f{=}2{+}1$ Clover by LHPC/BMW~\cite{Green:2012ej};
$N_f{=}2{+}1$ Domain Wall fermions (DWF) by LHPC~\cite{Green:2012ej};
$N_f{=}2{+}1$  DWF/staggered by LHPC~\cite{Green:2012ej};
$N_f{=}2{+}1{+}1$ Twisted Mass fermions (TMF) by ETMC~\cite{Alexandrou:2013jsa};
$N_f{=}2$ Clover fermions by RQCD~\cite{Bali:2014nma};
$N_f{=}2$ TMF $\&$ Clover by ETMC~\cite{Abdel-Rehim:2015owa};
$N_f{=}2{+}1{+}1$ Highly Improved Staggered Quarks (HISQ) by PNDME~\cite{Bhattacharya:2016zcn}.
As mentioned above, the scalar charge suffers from excited state contamination, so it is important to perform different
analysis to address this systematics. A high-statistics analysis was performed by ETMC at the physical pion mass
\cite{Abdel-Rehim:2015owa} which shows an increasing trend or $g^{u-d}_S$ as the source-sink separation increases, as presented in
the left plot of Fig.~\ref{fig_gS_ratio}. The plateau method, a two-state fit and the summation methods were applied, with the latter
resulting a much larger value for $g^{u-d}_S$. In the right plot of Fig.~\ref{fig_gS_ratio} we show $g^{u-d}_S$ from a recent work by PNDME
which includes an ensemble at the physical point~\cite{Bhattacharya:2016zcn}. The two-state fit lead to a larger estimate than
the plateau fits, which is compatible with the findings by ETMC.
\begin{figure}[!h]
\cl{\includegraphics[scale=0.22]{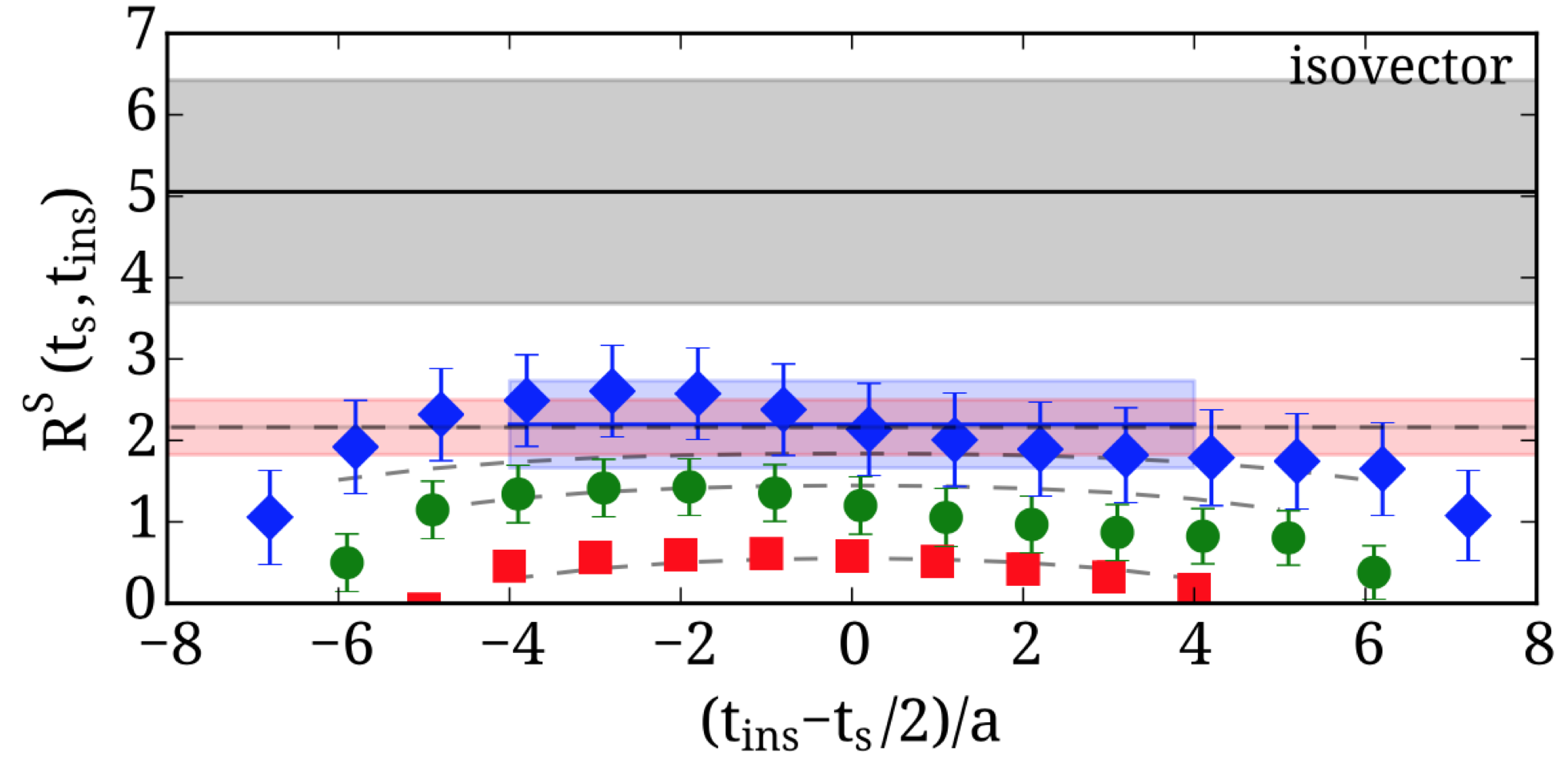}\,\, 
\includegraphics[scale=0.64]{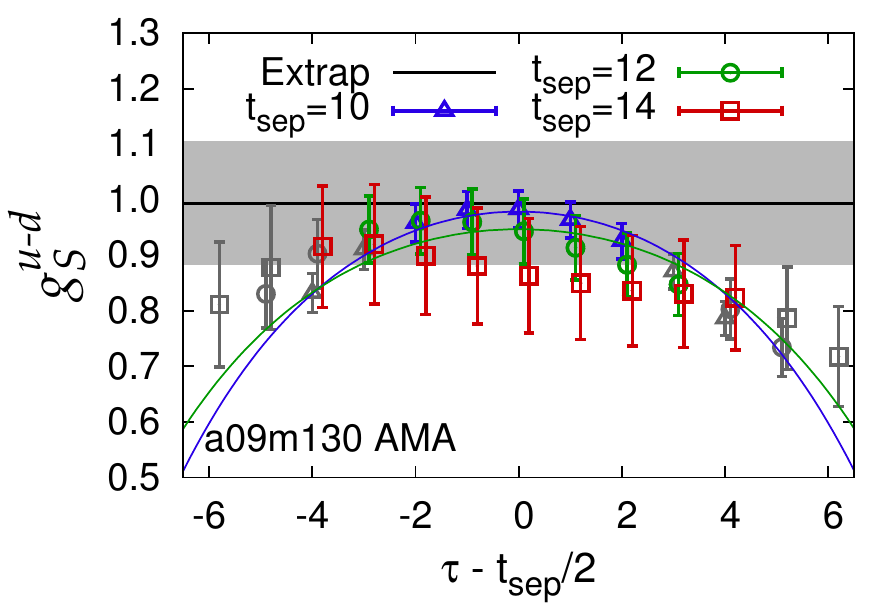}}
\caption{Ratio leading to $g^{u-d}_S$ as a function of the source-sink separation. The left plot corresponds to $N_f{=}2 $ TMF
\cite{Abdel-Rehim:2015owa} and the right plot to $N_f{=}2{+}1{+}1$ HISQ~\cite{Bhattacharya:2016zcn}. }
\label{fig_gS_ratio}
\end{figure}
\FloatBarrier
The pion mass dependence of the connected isoscalar combination for the scalar charge is demonstrated in Fig.~\ref{fig_gS_IS},
for a selection of lattice discretizations: $N_f{=}2{+}1$ Clover fermions, DWF, DWF/staggered (all from Ref.~\cite{Green:2012ej}), 
$N_f{=}2{+}1{+}1$ TMF~\cite{Alexandrou:2013jsa};
$N_f{=}2$ Clover fermions~\cite{Bali:2014nma};
$N_f{=}2$ TMF $\&$ Clover term~\cite{Abdel-Rehim:2015owa};
$N_f{=}2{+}1{+}1$ HISQ~\cite{Bhattacharya:2013ehc}.
The main observation is that $g^{u+d}_S$ exhibits strong pion mass dependence for all discretization, which implies a sensitivity to 
the fit functions for a chiral extrapolation. Thus, simulations at the physical point are crucial for the elimination of uncontrolled 
systematics related to such extrapolation.
\begin{figure}[!h]
\cl{\includegraphics[scale=0.32,angle=-90]{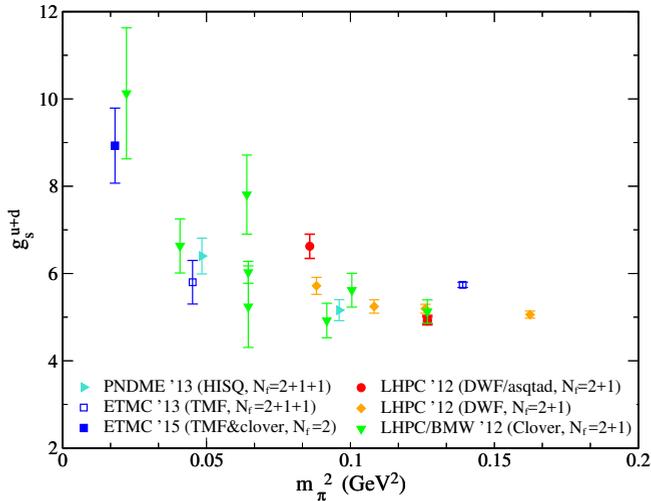}}
\caption{Lattice results for $g^{u+d}_S$ as a function of the $m_\pi^2$, corresponding to: 
$N_f{=}2{+}1$ Clover by LHPC/BMW~\cite{Green:2012ej} (down green triangles);
$N_f{=}2{+}1$ DWF by LHPC~\cite{Green:2012ej} (orange diamonds);
$N_f{=}2{+}1$  DWF/staggered by LHPC~\cite{Green:2012ej} (red circles);
$N_f{=}2{+}1{+}1$ TMF by ETMC~\cite{Alexandrou:2013jsa} (open blue square);
$N_f{=}2$ TMF $\&$ Clover by ETMC~\cite{Abdel-Rehim:2015owa} (filled blue square);
$N_f{=}2{+}1{+}1$ HISQ by PNDME~\cite{Bhattacharya:2013ehc} (turquoise right triangles).} 
\label{fig_gS_IS}
\end{figure}
\FloatBarrier

\subsubsection{Tensor Charge}
\label{sub3_1_2}

\begin{figure}[!h]
\cl{\includegraphics[scale=0.34, angle=-90]{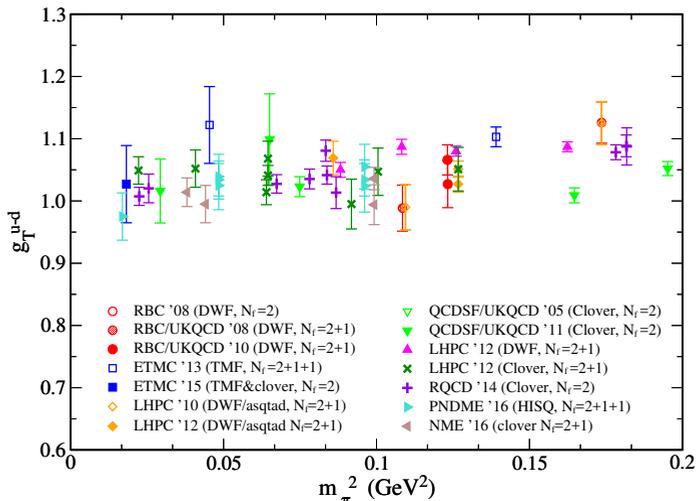}}
\caption{Lattice results for $g^{u-d}_T$ as a function of the $m_\pi^2$, corresponding to: 
$N_f{=}2$ Clover (QCDSF/UKQCD~\cite{Gockeler:2005cj,Pleiter:2011gw}, RQCD~\cite{Bali:2014nma}),
$N_f{=}2$ DWF (RBC~\cite{Lin:2008uz}),
$N_f{=}2{+}1$ DWF (RBC/UKQCD~\cite{Ohta:2008kd,Aoki:2010xg}, LHPC~\cite{Green:2012ej}),
$N_f{=}2{+}1$ DWF on asqtad sea (LHPC~\cite{Bratt:2010jn,Green:2012ej}), 
$N_f{=}2{+}1$ Clover (LHPC~\cite{Green:2012ej}),
$N_f{=}2{+}1{+}1$ TMF (ETMC~\cite{Alexandrou:2013wka}), 
$N_f{=}2{+}1{+}1$ HISQ (PNDME~\cite{Bhattacharya:2016zcn}), 
$N_f{=}2$ TMF with Clover (ETMC~\cite{Abdel-Rehim:2015owa}),
$N_f{=}2{+}1$ Clover (NME~\cite{Yoon:2016jzj}).} 
\label{fig_gT}
\end{figure}

\begin{figure}[!h]
\cl{\includegraphics[scale=0.2]{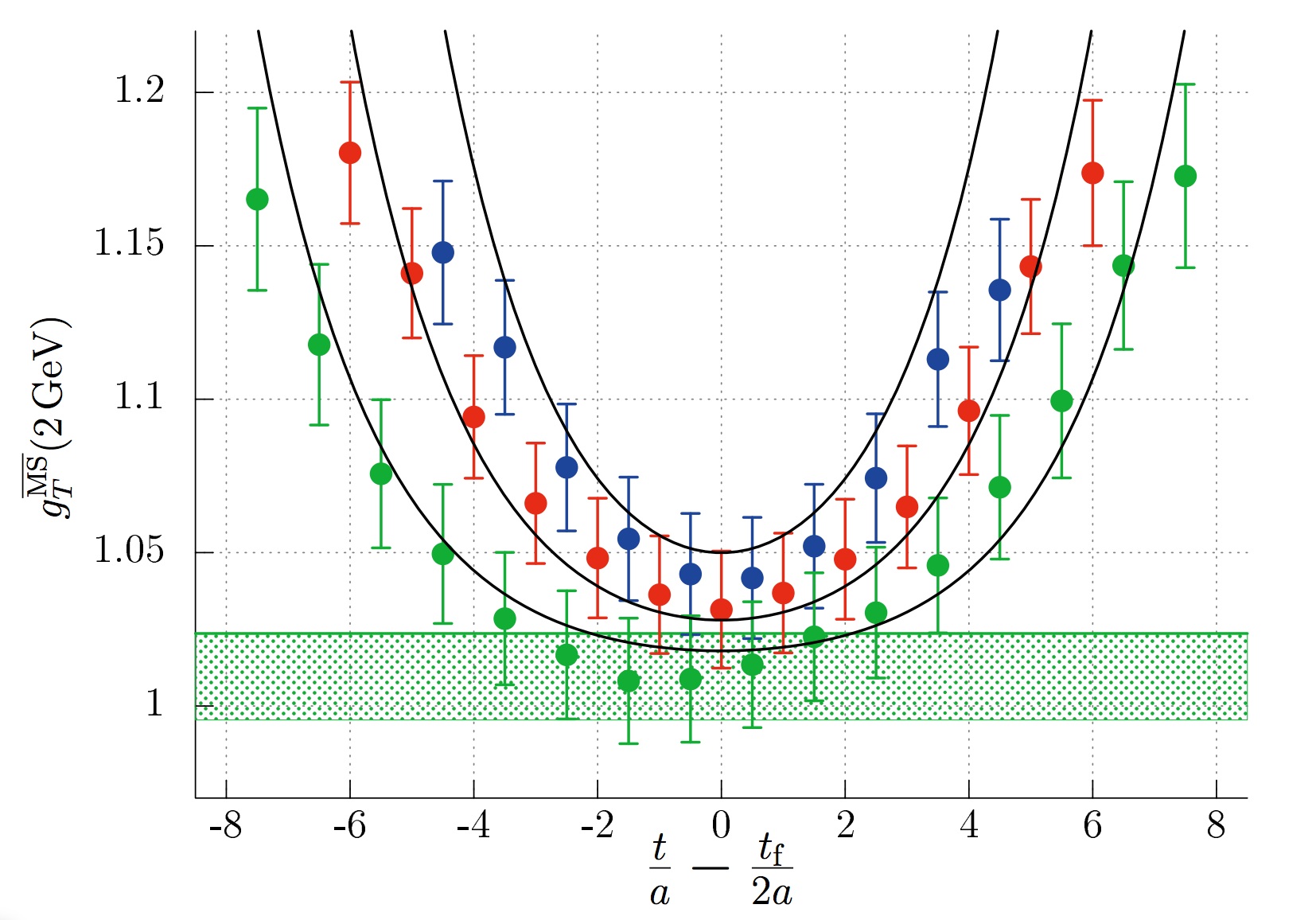}}
\caption{Ratio for $g^{u+d}_T$ as a function of the current insertion time, for source-sink separations $t_f{=}9a, 12a, 15a$. 
The shaded region indicates the fitted value of $g^{u-d}_T(2 GeV)$ and the corresponding statistical uncertainty.
The ensemble correspond to $N_f{=}2$ Clover fermions at $m_\pi{=}150$ MeV~\cite{Bali:2014nma}.}
\label{fig_gT_ratio}
\end{figure}

The computation of the tensor charge is quite timely since experiments using polarized 3He/Proton at Jefferson lab 
aim at increasing the experimental accuracy of its measurement by an order of magnitude~\cite{Gao:2010av}.  
The tensor charge, which is the zeroth moment of the transversity distribution functions, the last part among
the three quark distributions at leading twist
\be
\langle N(\vec{p'}) \, \mathcal{O}_X \, N(\vec{p}) \rangle \Big{|}_{Q^2=0} \,, \quad 
\mathcal{O}^\alpha_X = {\cal O}_T^{\alpha\,\mu\nu} = \bar{\psi}\,\sigma^{\mu\nu} \frac{\tau^\alpha}{2}\,\psi\,.
\ee
\noindent
Results on the isovector tensor charge are compared in Fig.~\ref{fig_gT} 
for various discretizations, number of dynamical quarks, lattice volumes and lattice spacings. For a meaningful comparison, 
we plot results extracted from the plateau method. All data exhibit mild pion mass dependence and overall there is a 
very good agreement among lattice data, with very small sensitivity on the action parameters.
We would like to highlight the data close or at the physical point which correspond to 
$N_f{=}2$ Clover fermions~\cite{Pleiter:2011gw,Bali:2014nma},
$N_f{=}2{+}1$ Clover fermions~\cite{Green:2012ej},
$N_f{=}2$ TMF $\&$ Clover~\cite{Abdel-Rehim:2015owa}, and
$N_f{=}2{+}1{+}1$ HISQ (PNDME~\cite{Bhattacharya:2016zcn}).
\FloatBarrier
The excited states contamination for $g^{u-d}_T$ have been also investigated, 
revealing a weak dependence on the source-sink time separation
\cite{Bali:2014nma,Bhattacharya:2016zcn,Abdel-Rehim:2015owa}. As an example
we demonstrate the work by RQCD~\cite{Bali:2014nma} in Fig.~\ref{fig_gT_ratio}
corresponding to $N_f{=}2$ Clover fermions at $m_\pi{=}150$ MeV. As seen from the 
plot the values from the plateau method are consistent for separation of $9a$ to $15a$ 
and are in agreement with the value extracted from the two-state fit (shaded area), within 
statistical uncertainties.

Experimentally, there exist data for $g_T$ obtained from combined global analysis of the measured azimuthal asymmetries in
SIDIS and in $e^+ e^- \to h_1 h_2 X$ (see, e.g. \cite{Anselmino:2013vqa,Radici:2015mwa}).
There are also results from model predictions, but direct comparison of the tensor charges from different models and scales is
not always meaningful, since the tensor charges are strongly scale-dependent quantities. A recent value for the isovector tensor
charge is $g^{u-d}_T{=}0.81(44)$ at 2GeV, which has a very large uncertainty compared to the LQCD results discussed above.

\FloatBarrier

\vspace*{-0.5cm}
\subsection{Neutron Electric Dipole Moment}
\label{sub3_2}

Recently there has been a lot of interest in the neutron electric dipole moment (nEDM)
both experimentally and theoretically. In this section we review recent lattice calculations for the nEDM,
$\vec{d}_N$, using different techniques. The EDM is a probe of Physics BSM, as a non-zero value would indicate the violation of parity
($P$) and time ($T$) symmetries, and consequently of $CP$~\cite{Pospelov:2005pr}. So far, no finite neutron EDM (nEDM) has been
reported and although there are current bounds they are several orders of magnitude
below what one expects from $CP$-violation induced by weak interactions. To date, the best experimental upper limit 
is~\cite{Helaine:2014ona,Baker:2006ts,Baker:2007df}
\be
\vert \vec{d}_N \vert  < 2.9 \times 10^{-13} e \cdot {\rm fm} \ (90\% \ {\rm CL})\,.
\label{eq:nEDM_smaller_experimental_upperbound}
\ee 
Thus, a precise determination of the nEDM from first principles may provide a valuable input for future experiments, which is one
of the motivations of such lattice calculations. Theoretically an EDM may arise by adding in the $CP$-conserving QCD Lagrangian density 
a $CP$-violating interaction term, proportional to the topological charge, $q$
\bea
  {\cal L}_{\rm QCD} \left( x \right) = \frac{1}{2 g^2} 
{\rm Tr} \left[ G_{\mu \nu} \left( x \right) G_{\mu \nu} \left( x \right) \right] + 
\sum_{f} {\overline \psi}_{f} \left( x \right) (\gamma_{\mu} D_{\mu} +m_f) \psi_{f}\left( x \right)\,
- i \theta q \left( x \right)\,,\\
  q \left( x \right) = \frac{1}{ 32 \pi^2} \epsilon_{\mu \nu \rho \sigma} 
{\rm Tr} \left[ G_{\mu \nu} \left( x \right) G_{\rho \sigma} \left( x \right) \right]\,,\hspace*{3cm}
\eea
where $\theta$ is the parameter controlling the strength of the $CP$-breaking, and can be taken as a small continuous
parameter allowing a perturbative expansion and only keep first order contributions in $\theta$. This is in accordance to effective
field theory calculations (see references within~\cite{Alexandrou:2015spa})
that give a bound of the order 
$\theta \lesssim {\cal O} \left(10^{-10} - 10^{-11} \right)$.
At leading order of  $\theta$ and in Euclidean space, nEDM is extracted from a
the zero momentum transfer limit of the $CP$-odd form factors that appears in the
decomposition of the nucleon matrix element using a vector current, at the presence of the
$CP$-odd term in the Lagrangian, that is
\be
\vert \vec{d}_N \vert =  \theta \lim_{Q^2 \to 0} \frac{\vert F_3(Q^2) \vert}{2 m_N}\,.
\label{eq:dN}
\ee
In the above equation $m_N$ is the neutron mass, $Q^2$ the four-momentum transfer in Euclidean space and
$F_3(Q^2)$ is the $CP$-odd form factor. In the decomposition of the 
$CP$-violating matrix element the $F_3(Q^2)$ comes with a factor of the momentum ($Q_k F_3(Q^2)$), hindering a
direct evaluation of $F_3(0)$. There are techniques developed for removing the unwanted factor of the momentum, based
on a method similar to a continuum derivative, as well as, a momentum-elimination technique. Details on different methods and
for the extraction of $F_3(Q^2)$ can be found in Refs.~\cite{Alexandrou:2016rbj,Alexandrou:2015spa}.
Similar methods have been extended for the calculation of other matrix elements~\cite{Bouchard:2016gmc}.

\begin{figure}[!h]
\cl{\includegraphics[scale=0.22]{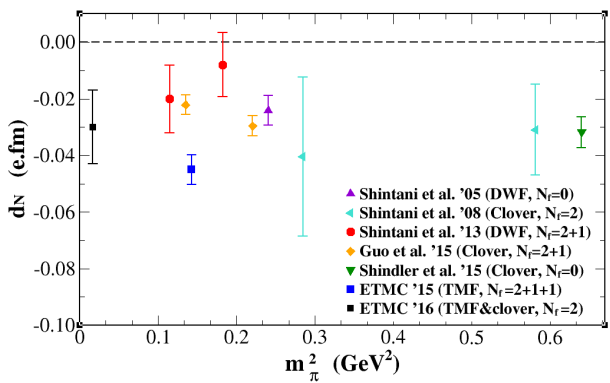}}
\caption{The nEDM versus $m_\pi^2$ for: a) $N_f{=}2{+}1{+}1$ TMF
\cite{Alexandrou:2015spa} (blue square) corresponding to a
weighted average using different methods for extracting $F_3(0)$), 
b) $N_f{=}0$ DWF~\cite{Shintani:2005xg} (magenta upward triangles),
$N_f{=}2{+}1$ DWF~\cite{Shintani:2014zra} (red circles) and 
$N_f{=}0$ Clover fermions~\cite{Shindler:2015aqa} (green downward triangles) by extracting
the $CP$-odd $F_3(Q^2)$ and fitting its $Q^2$-dependence, c) $N_f{=}2$
Clover fermions~\cite{Shintani:2008nt} (turquoise left triangles) 
obtained using a background electric field, d) $N_f{=}2{+}1$ Clover
fermions~\cite{Guo:2015tla} (orange diamonds) by implementing an
imaginary $\theta$.}
\label{nEDM}
\end{figure}

So far we discussed the extraction of the nEDM from the $CP$-odd form factor
\cite{Shintani:2005xg,Shintani:2014zra,Shindler:2015aqa,Alexandrou:2015spa}, which is
based on the small-$\theta$ expansion. However, there are alternative methods to 
compute the nEDM, such as, simulating the action with an imaginary $\theta$~\cite{Guo:2015tla}, 
or with an application of an external electric field and measuring the associated
energy shifts \cite{Shintani:2008nt}. Lattice results extracted from all the
aforementioned methods are collected in Fig.~\ref{nEDM}, where each work employs 
different definition of the topological charge. The black square at the physical point is a preliminary
result for the nEDM using TMF $\&$ Clover by ETMC. The large statistical error is a demonstration
of the difficulty to extract this quantity and the need for development of new techniques. 

\FloatBarrier

\vspace*{-0.5cm}
\section{Summary}
\label{sec4}

Understanding nucleon structure from first principles is considered a milestone of hadronic
Physics and numerous theoretical investigations have been devoted to its study. In these proceedings
we have presented recent developments in nucleon structure using LQCD, related to new Physics searches.
Such quantities are the nucleon scalar and tensor charges and the neutron electric dipole moment, presented
here. 

There have been major advances in algorithms and techniques, as well as, increase in the computational power, that allowed 
simulations with the quark masses fixed to their physical values. These simulations at the physical point have eliminated one 
of the systematic uncertainties that was inherent in all lattice calculations in the past, that is, the difficulty to quantify 
systematic error due to the chiral extrapolation.

Although there is significant progress in LQCD calculations, there are several challenges: development of noise-reduction 
algorithms at low cost and addressing systematic uncertainties in order to compute reliably quantities 
that reproduce experimental data or can probe beyond the Standard Model Physics.

\bigskip
{\bf{Acknowledgements}}:
 I would like to thank all members of the ETMC for a fruitful collaboration and in particular 
 C. Alexandrou, A. Abdel-Rehim, K. Jansen, K. Hadjiyiannakou, Ch. Kallidonis,
G. Koutsou, H. Panagopoulos and A. Vaquero for their invaluable contributions.

%
\bibliography{references}

\end{document}